\documentclass[10pt, conference, letterpaper]{ IEEEtran}
\IEEEoverridecommandlockouts
\usepackage{cite}
\usepackage{amsmath,amssymb,amsfonts}
\usepackage{algorithmic}
\usepackage{graphicx}
\usepackage{textcomp}
\usepackage{xcolor}
\usepackage{bm}
\usepackage[utf8]{inputenc}
\usepackage{booktabs}
\def\BibTeX{{\rm B\kern-.05em{\sc i\kern-.025em b}\kern-.08em
    T\kern-.1667em\lower.7ex\hbox{E}\kern-.125emX}}
\begin{document}

\title{RadCloudSplat: Scatterer-Driven 3D Gaussian Splatting with Point-Cloud Priors for  Radiomap Extrapolation \thanks{The work was supported in part by the National Key Research and Development Program of China under Grant 2024YFA1014202; in part by the National Natural Science Foundation of China under Grant 62301334, Grant 62031008, Grant 62231019 and Grant U23B2005; in part by the Shenzhen Science and Technology Program under Grant RCIC20210609104448114 and Grant ZDSYS20230626091302006; in part by Hetao Shenzhen-Hong Kong Science and Technology Innovation Cooperation Zone Project (No. HZOSWSKCCYB-2024016); in part by the Guangdong Major Project of Basic and Applied Basic Research (No. 2023B0303000001); in part by the SRIBD Internal Project J00120250004.  {\em (Corresponding author: Ye Xue)}\\The authors have provided public access to their code at https://github.com/yeehengwang/RadCloudSplat.}
}
\author{
\IEEEauthorblockN{Yiheng Wang$^{1,2,3}$, Ye Xue$^{1\#}$, Shutao Zhang$^{4}$, Hongmiao Fan$^{4}$, Tsung-Hui Chang$^{2,3}$}\\
\IEEEauthorblockA{$^{1}$School of Intelligent Systems Engineering, Shenzhen Campus of Sun Yat Sen University, Guangdong, China\\
$^{2}$Shenzhen Research Institute of Big Data, The Chinese University of Hong Kong-Shenzhen, Guangdong, China\\
$^{3}$School of Artificial Intelligence, The Chinese University of Hong Kong-Shenzhen, Guangdong, China\\
$^{4}$Networking and User Experience Laboratory, Huawei Technologies,  Shenzhen, Guangdong, China\\
}
}

\maketitle

\begin{abstract}
A radiomap represents the spatial distribution of wireless signal strength, which is critical for applications like network optimization. However, constructing a radiomap relies on measuring radio signal power across the entire system, which is costly in outdoor environments due to large network scales. We present RadCloudSplat, a framework that extends 3D Gaussian Splatting (3DGS) to radio frequencies for efficient and accurate radiomap extrapolation from sparse measurements. RadCloudSplat models environmental scatterers and radio paths using 3D Gaussians, capturing key factors of radio wave propagation. It employs a relaxed-mean (RM) scheme to reparameterize the positions of 3D Gaussians from noisy and dense 3D point clouds. A camera-free 3DGS-based projection is proposed to map 3D Gaussians onto 2D radio beam patterns. Furthermore, a regularized loss function and recursive fine-tuning using highly structured sparse measurements in real-world settings are applied to ensure robust generalization. Experiments on synthetic and real-world data show state-of-the-art extrapolation accuracy and execution speed, solidifying the framework’s credibility for real-world deployment.
\end{abstract}

\section{Introduction}

As the demand for seamless connectivity in next-generation wireless networks continues to rise, real-world outdoor cellular network deployments face persistent challenges such as weak signal strength, limited communication bandwidth, suboptimal infrastructure placement, and vulnerabilities in communication security and privacy. These issues can significantly degrade the reliability and performance of wireless connections. To mitigate such challenges, {\em radiomaps}, which capture the spatial distribution of concurrent wireless transmissions as a function of geographic locations, are widely adopted as a standard tool to evaluate since they can offer critical insights into {\em beamspace Received Signal Strength} (RSS) patterns and radio-frequency (RF) activity across outdoor environments, thereby enabling a range of key applications including network optimization\cite{wang2025gnn}, resource allocation\cite{bi2019engineering}, and autonomous driving~\cite{chowdappa2018distributed}.

However, constructing the radiomap with dense signal strength measurements is highly impractical, particularly in outdoor environments. Signal strength data are typically sparse, as they are collected manually through labor-intensive methods such as drive tests, where engineers traverse coverage areas with specialized equipment to sample wireless signals at specific locations. This process is costly and limited in scope, resulting in incomplete radiomaps. Moreover, outdoor environments are vast and complex, and wireless signals are influenced by obstacles, interference, and environmental conditions, making robust and efficient radiomap reconstruction methods essential.

Traditional approaches for radiomap reconstruction, such as \textit{inverse distance weighting (IDW)}~\cite{kuo2010discriminant} and \textit{Kriging}~\cite{sato2020radio}, rely on fixed-function models to interpolate missing data. While computationally efficient, these methods struggle to handle the complexity of outdoor environments and fail to capture the underlying physics of wireless signal propagation. \textit{Ray Tracing}~\cite{10155734} methods simulate wireless signal paths using computer-aided design (CAD) representations of 3D environments, but such models often fail to adapt to real-world scenarios due to discrepancies between modeled and actual 3D environments. Learning-based methods, such as \textit{variational autoencoders (VAE)}~\cite{teganya2021deep} and \textit{deep convolutional GANs (DCGAN)}~\cite{radford2015unsupervised}, offer data-driven alternatives by avoiding explicit assumptions about signal propagation. However, these approaches require dense measurements for effective training and fail to incorporate the physics of wireless signals, limiting their utility in sparse, real-world settings.

 One of the promising advancements in learning-based models is the recent adoption of the neural radiance field (NeRF),  a breakthrough in computer vision for view synthesis\cite{mildenhall2021nerf}, such as $\text{NeRF}^2$\cite{zhao2023nerf2} and NeWRF\cite{lunewrf}, which have shown promising results in radiomap reconstruction by modeling the spatial distribution of wireless signals. However, these methods are computationally intensive and exhibit slow synthesis speeds, rendering them impractical for large-scale outdoor environments where efficiency and scalability are paramount. In contrast, 3D Gaussian Splatting (3DGS)\cite{kerbl20233d} has recently emerged as a powerful alternative in the fields of computer vision and graphics, offering real-time performance and efficient scene representation and driving substantial progress \cite{zhou2024drivinggaussian,li2024dngaussian,10655197}. Despite these innovations, most applications have predominantly focused on the visible light spectrum. A notable exception is WRF-GS\cite{wen2024wrf}, which adapted 3DGS for wireless channel modeling. However, WRF-GS depends on assumptions specific to simplified channel models that may not generalize well to real-world scenarios. More critically, it relies solely on multipath profile data~\cite{zhao2023nerf2}, overlooking the {\em beamspace} signal representations and limiting its use to constrained indoor environments.


To address these limitations, we present the first extension of 3DGS to the radio frequency domain using \textit{beamspace} RSS data collected from real-world outdoor cellular networks. This approach enables robust and spatially-aware modeling of wireless signal radiation patterns in the form of {\em radiomaps}, making it particularly suitable for practical MIMO-based deployments~\cite{wang2024neural}, where multi-path propagation and multi-beam effects are present. Meanwhile, 3DGS-based models are computationally efficient and align well with the propagation characteristics of wireless signals for the outdoor environment. Their anisotropic 3D Gaussian volumetric representation is suited to modeling radio scatterers that influence signal behavior. Furthermore, LiDAR point clouds, providing accurate representations of outdoor environments~\cite{wang20243d}, can be directly utilized to initialize 3DGS, enabling these models to effectively capture interactions between signals and the environment~\cite{sun2023define}. However, applying 3DGS to radiomap reconstruction presents unique challenges. First, processing dense point cloud data in outdoor large-scale environments incurs substantial computational overhead~\cite{mi2024measurement}. Also, point cloud data are often affected by noise, which leads to inaccurate or unreliable representations of the environment. Second, traditional 3DGS relies on camera-based calibration to map 3D world coordinates to 2D projections, which is infeasible for wireless signals. Finally, sparse signal strength measurements lack sufficient data for accurate extrapolation in uncovered regions, making robust generalization a critical challenge.

To address these challenges, we propose \textit{RadCloudSplat}, a framework that extends 3D Gaussian Splatting to the radio frequency domain for efficient and accurate radiomap reconstruction. Our contributions are summarized as follows:
\begin{itemize}
\item \textbf{Relaxed-Mean (RM) reparameterization:} We propose a novel RM scheme to select key virtual scatterers from noisy and dense point clouds, providing an accurate reparameterization of means for radio 3D Gaussians. These novel trainable parameters in RM are jointly optimized with the whole model's other parameters, enabling efficient gradient-based training.
    \item \textbf{Camera-free 3DGS framework:} We introduce the first camera-free 3DGS-based framework for wireless signal strength reconstruction, integrating radio multi-path propagation characteristics and environmental semantics. This framework learns wireless radiation patterns from sparse signal strength measurements within milliseconds, even in complex outdoor environments.  
    \item \textbf{Extrapolation-enhancing techniques:} To improve extrapolation performance, we propose a ternary regularized loss function and a recursive fine-tuning scheme. The fine-tuning process leverages predicted data to iteratively refine the model for improved accuracy in regions with sparse or no measurements.
\end{itemize}

\section{Preliminaries}\label{sec:prelim}

In this section, we present preliminaries on wireless signal propagation and 3D Gaussian Splatting, focusing on how RSS data are collected and the challenges of radiomap reconstruction.

\subsection{Wireless Signal Propagation Mechanism \label{sec:radioprop}} 
\begin{figure}[ht]
\vspace{-2mm}
\begin{center}
\centerline{\includegraphics[width=0.8\columnwidth]{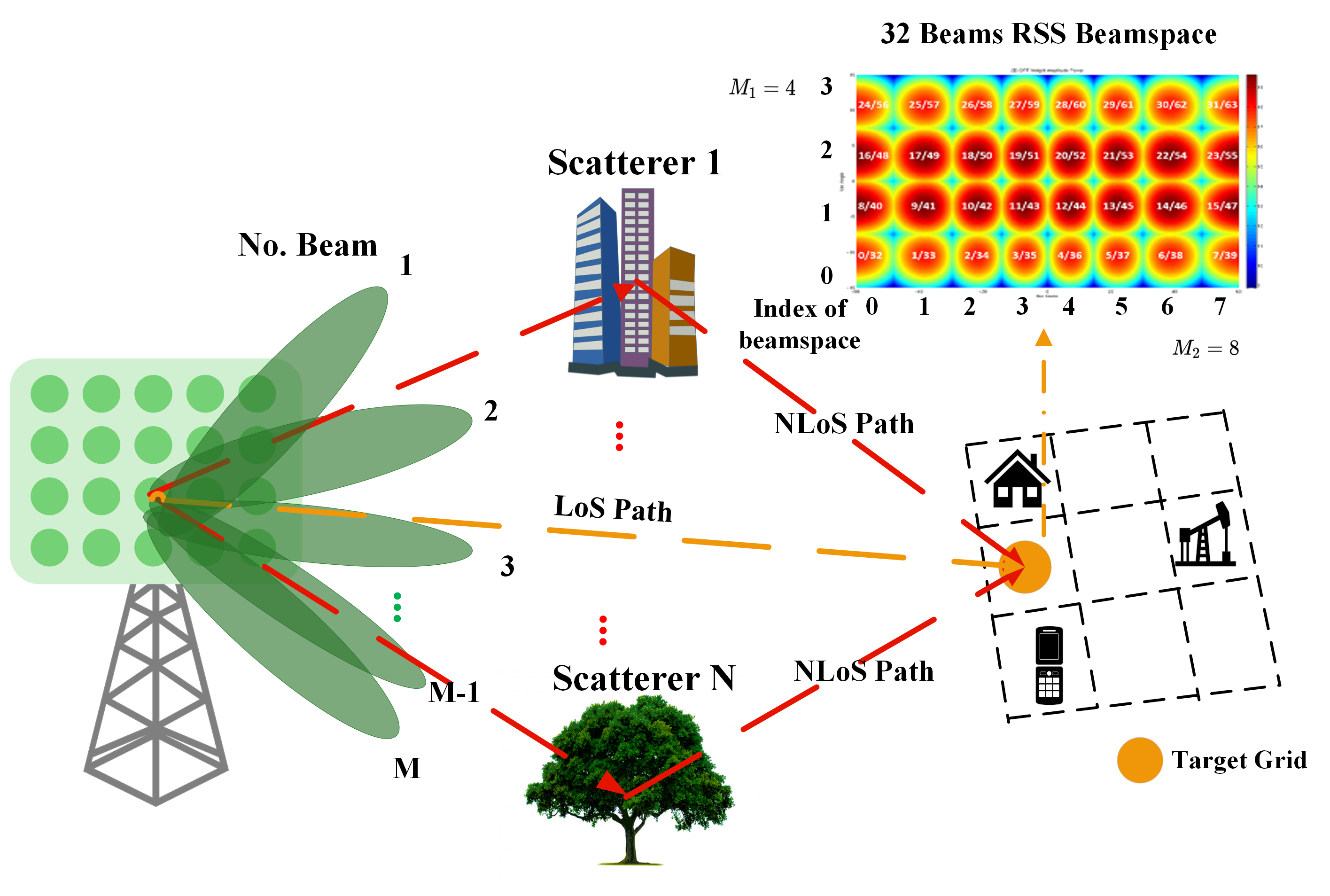}}
\caption{Illustration of wireless communication setup and RSS {\em beamspace}.}
\label{fig:System Model}
\end{center}
\vspace{-5mm}
\end{figure}

Radiomaps are 2D representations of the spatial distribution of RSS, which quantifies the power of wireless signals across a region. In modern wireless communication systems, RSS is often measured in the {\em beamspace}. {\em Beamspace} represents wireless signals based on beam directions using beamforming techniques\cite{ning2023learning}. Unlike omnidirectional transmission, beamforming concentrates signals into specific beams, where the predefined beams divide the coverage space into angular regions, as shown in Fig.~\ref{fig:System Model}. The beamspace representation organizes RSS according to predefined $M=M_1\times M_2$ beam angles. For the $m$-th beam at the $l$-th location, the RSS can be computed as
\begin{equation}
\text{RSS}_{m,l} = P |\bm h_{m,l}|^2, \label{eq:rss}
\end{equation}
where $P$ is the transmission power and $\bm h_{m,l}$ denotes the wireless channel, which characterizes the effects of the propagation process, including signal attenuation and phase changes.

Wireless signal propagation is influenced by both a direct line-of-sight (LoS) path and multiple non-line-of-sight (NLoS) paths caused by environmental scatterers such as buildings and walls. These scatterers induce phenomena such as reflection, diffraction, and scattering \cite{kapusuz2014determination}, resulting in a phenomenon known as {\em multi-path propagation} \cite{bertoni1999radio}. {\em Multi-path propagation} significantly impacts RSS, as the received signal is the superposition of multiple propagation paths, each contributing distinct attenuation and phase changes. The wireless channel, which encapsulates the impact of the propagation process at location $l$, aggregating all $M$ beams, can be expressed as
\begin{equation}
\label{discretizing}
\bm{h}_l = \sum_{n=1}^{N} \Delta\bm{\beta}_n \odot e^{j\Delta\bm{\varphi}_n},
\end{equation}
where $N$ is the total number of paths caused by $N$ scatterers\footnote{Here we assume each scatter induces one radio propagation path for simplicity}, $\Delta\bm{\beta}_n\in\mathbb{R}^{M}$ denotes the attenuation of the signal strength, and $\Delta\bm{\varphi}_n\in\mathbb{R}^{M}$ represents the phase shift contributed by the $n$-th path. $\odot$ denotes the Hadamard product.

The inherent complexity of multi-path propagation, combined with the variability of wireless environments, poses significant challenges for constructing accurate radiomaps. Sparse RSS measurements further exacerbate these difficulties, as manual field surveys (e.g., drive tests) are labor-intensive and costly, leaving large terrain areas undersampled. These limitations lead to incomplete or low-resolution radiomaps, which fail to capture the intricate spatial variations in signal power caused by environmental factors. 

\subsection{3D Gaussian Splatting (3DGS)\label{sec:3dgsback}} 3DGS \cite{kerbl20233d} is an explicit radiance field-based scene representation that models a radiance field using a large number of total $I$ 3D anisotropic balls, and the $i$-th 3D Gaussian is defined as
\begin{align}
    G_i(\bm{x}) = e^{-\frac{1}{2}(\bm{x} - \bm{\mu}_i)^\textnormal{T} {\bf \Sigma}^{-1}_i (\bm{x} - \bm{\mu}_i)},
\end{align}
where $\bm{x} = (x_0, x_1, x_2)$ denotes the three-dimensional spatial coordinates of the Gaussian point and $\bm{\mu}_i \in \mathbb{R}^3$ and ${\bf \Sigma}_i\in \mathbb{R}^{3\times3}$ denote the mean and variance of the Gaussian representation\footnote{The covariance matrix ${\bf \Sigma}_i$, which defines the ellipsoid shape of the $i$-th Gaussian, can be expressed in terms of a scaling matrix ${\bf S}_i \in \mathbb{R}^{3 \times 3}$ and a rotation matrix ${\bf R}_i \in \mathbb{R}^{3 \times 3}$ using its Eigen Decomposition. Here, the rotation matrix ${\bf R}_i$ is parameterized using quaternions $\bm{r}_i \in \mathbb{R}^4$, while the scaling factors are represented by $\bm{s}_i \in \mathbb{R}^3$.}. Besides, the $i$-th anisotropic ball is also characterized by parameters that include its opacity $\alpha_i \in \mathbb{R}$ and the Spherical Harmonics (SH) coefficients $\bm{\tau}_i \in \mathbb{R}^{C \times (k+1)^2}$ (where $k$ is the degree of the SH function $\bm{sh}(\cdot, k)$) \cite{fridovich2022plenoxels} for modeling view-dependent color. {Specifically, the color of the $i$-th 3D Gaussian at the view direction $\bm v_i$ is
\begin{align}
    \bm{c}_i= \bm{sh}(\bm{\tau}_i,\bm v_{i}, k) \in \mathbb{R}^C,
\end{align}
where $C=3$ corresponds to the number of RGB channels, and the output dimension of the SH function depends on the first dimension of the SH coefficients.

{During the rendering phase, all 3D Gaussians $\{G_i(\bm{x})\}_{i=1}^{I}$
are projected (rasterized) onto the image plane, forming 2D Gaussian splats, which can be expressed as  $\{\bm {w}_i({p})=\text{Proj}_{\text{EWA}}(G_i(\bm{x}))\}_{i=1}^I$ at the ${p}$-th pixel, as described in \cite{zwicker2002ewa}. The 2D Gaussian splats are sorted from front to back tile-wisely, and \textit{alpha}-blending  algorithm is applied to approximate the $p$-th pixel RGB value $\bm{C}({p}) \in \mathbb{R}^3$ as 
\begin{align}
    \bm{C}({p}) = \sum_{i\in\mathbb I_p} \bm{c}_i \alpha_i\cdot\bm {w}_i({p})\prod_{j=1}^{i-1} (1 - \alpha_j\cdot\bm {w}_j({p})),
    \label{eq:alpha}
\end{align}
where $\mathbb I_p$ is the set of splats that contribute to the $p$-th pixel. After computing the color for each pixel, the rendered image is compared to the ground-truth image to calculate the pixel-wise loss. This loss is then used to update the model's properties parameters $\bm{\Phi} = \{(\bm{\mu}_i, \bm{\Sigma}_i,\alpha_i, \bm {\tau}_i )\}_{i=1}^{I}$.

\section{RadCloudSplat Framework}
{In this section, we introduce RadCloudSplat, a novel framework for RSS radiomap extrapolation in outdoor environments, which extends 3DGS to the radio frequency domain.}

\subsection{Problem Description and Framework Overview}\label{sec:problem}

\begin{figure}
\vspace{-5mm}
\begin{center}
\centerline{\includegraphics[width=0.5\columnwidth]{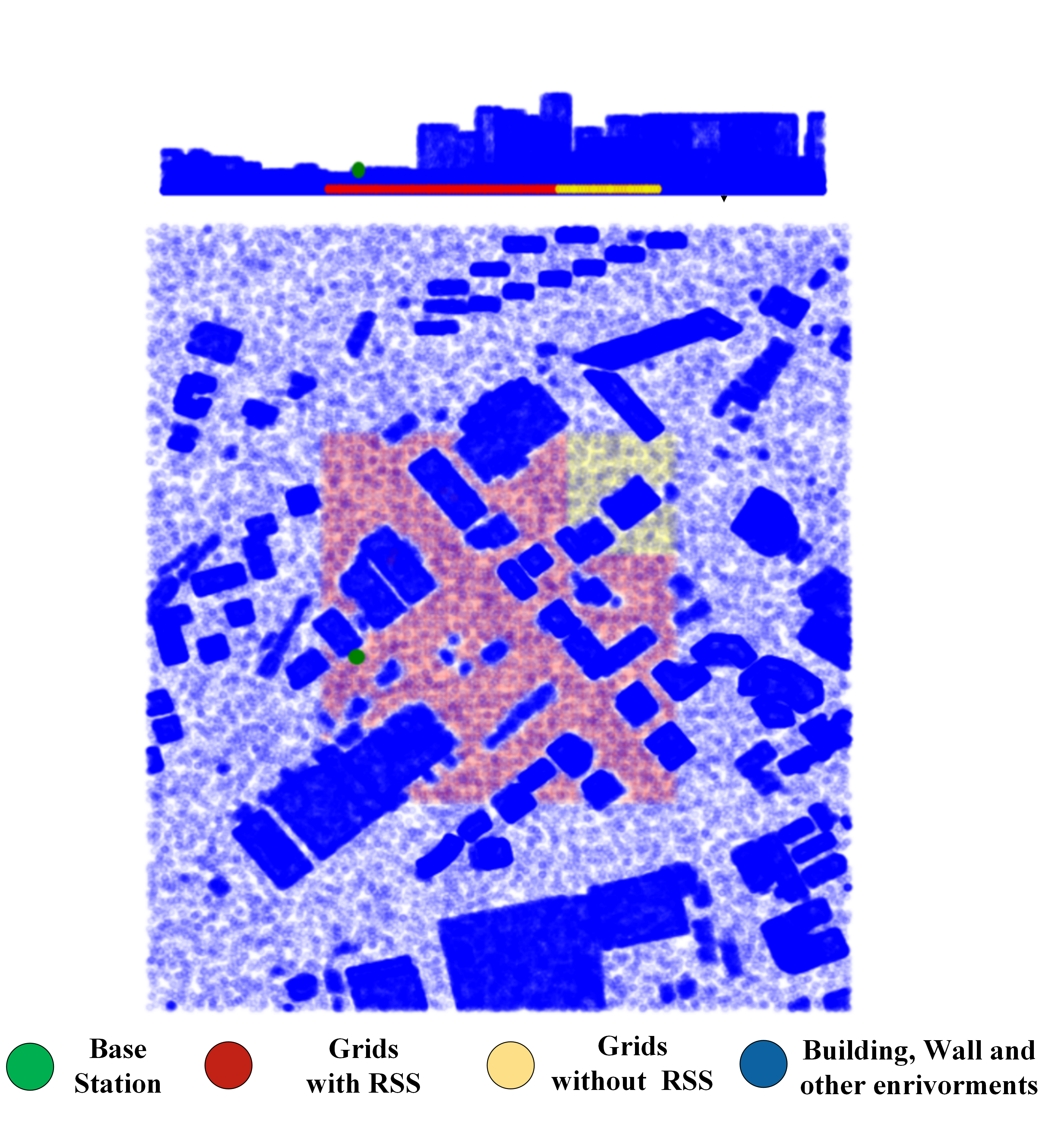}}
\caption{Illustration of the front and top views of a 3D point cloud of City A's outdoor environment.}
\label{fig:Task}
\end{center}
\vspace{-8mm}
\end{figure}

{Radiomap extrapolation in large-scale outdoor environments is challenging due to sparse and incomplete RSS measurements, which are typically collected only in limited regions (red grids $\mathbb L_k$ in Fig.~\ref{fig:Task}). This leaves large areas (yellow grids $\mathbb L_o$) without coverage, necessitating a method to predict RSS values in these unseen regions. To address this, RadCloudSplat aims to exploit environmental features, such as buildings and walls that strongly influence wireless signal propagation through reflection, scattering, and diffraction. However, obtaining such environmental features is infeasible in classical wireless measurement campaigns, as no scene details can be captured. Thanks to advancements in mobile terminals, such as intelligent vehicles and UAVs, these platforms \cite{DeepSense} can now capture 3D point cloud data, which can offer detailed geometric representations of the surrounding environment, enabling the RadCloudSplat to incorporate the spatial structure of obstacles that influence wireless signal propagation.

In this work, we propose \textbf{RadCloudSplat}, a novel framework that extends 3DGS to the radio frequency domain. RadCloudSplat leverages sparse RSS data, the position of BS, and environmental features derived from 3D point clouds to predict RSS values in complex, unseen locations. $N$ radio scatterers, such as buildings and walls, are represented using anisotropic 3D Gaussians, which capture their signal-related properties. These scatterers serve as the foundation for modeling signal propagation paths, and their characteristics are described as \textbf{RadCloudSplat attributes}
\begin{equation}
     \bm{\Phi}_{\text{rad}}\triangleq\{(\bm{\mu}_n^{\text{rad}}, \bm{\Sigma}_n^{\text{rad}}, \bm{\alpha}_n^{\text{rad}}, \bm{\tau}_n^{\text{rad}})\}_{n=1}^N,
\end{equation}
where $\bm{\mu}_n^{\text{rad}}$ represents the spatial location of the $n$-th scatterer in the environment, $\bm{\alpha}_n^{\text{rad}}$ is the signal attenuation reflecting both the magnitude attenuation and phase change of the radio signal caused by the $n$-th scatterer, $\bm{\Sigma}_n^{\text{rad}}$ is the covariance matrix that defines the anisotropic spatial uncertainty of the $n$-th scatterer, and $\bm{\tau}_n^{\text{rad}}$ is the RSS-encoded SH coefficients capturing how the scatterer’s contribution changes with respect to the direction of the signal propagation. 

Specifically, RadCloudSplat predicts RSS values $\{\hat{\bm{y}}_l \in \mathbb{R}^M\}$ for target locations ${P}_l \in \mathbb{R}^3$ given the BS position ${P}_{\text{BS}} \in \mathbb{R}^3$. This mapping is achieved using the RadCloudSplat attributes, collectively denoted as $\Phi_{\text{rad}}$. The prediction process is mathematically formalized as
\begin{equation}
    \label{eq:3dgsRSS}
    \mathcal{F}_{\bm{\Phi}_{\text{rad}}} : ({P}_{\text{BS}}, {P}_l) \to \hat{\bm{y}}_l.
\end{equation} To facilitate this process, RadCloudSplat employs a preprocessing model $\mathcal G$ that extracts key environmental information from dense and noisy 3D point clouds. Specifically, $\mathcal G$ identifies the positions of the key scatterers $\{\bm{\mu}_n^{\text{rad}}\}_{n=1}^N$, which form one of the critical RadCloudSplat attributes in $\bm{\Phi}_{\text{rad}}$. These scatterers' positions are essential for modeling the effects of signal propagation. However, extracting relevant scatterers from raw point cloud data is particularly challenging, given the sheer volume of data and the inherent noise in the measurements. To address this, RadCloudSplat incorporates an optimization-based preprocessing method, referred to as the \textbf{relaxed-mean (RM) scheme}, which reparameterizes the positions of the 3D Gaussians. Given the raw point cloud data $\hat{{P}}_{\text{raw}} \in \mathbb{R}^{R \times 3}$, consisting of $R$ points, and the BS position ${P}_{\text{BS}} \in \mathbb{R}^3$, the RM model is formulated as
\begin{equation}
    \mathcal{G}: ({P}_{\text{BS}}, \hat{{P}}_{\text{raw}}) \to \{\bm{\mu}_n^{\text{rad}}\}_{n=1}^N.
\end{equation}

The RM scheme will be elaborated in Section ~\ref{sec:relax}. Once the scatterer positions are identified, additional RadCloudSplat attributes are optimized to ensure accurate RSS predictions. The complete set of attributes and the rendering process will be discussed in Section ~\ref{sec:Synthesis}.

\begin{figure*}[t]
\centering
   \centering
\includegraphics[width=0.8\textwidth]{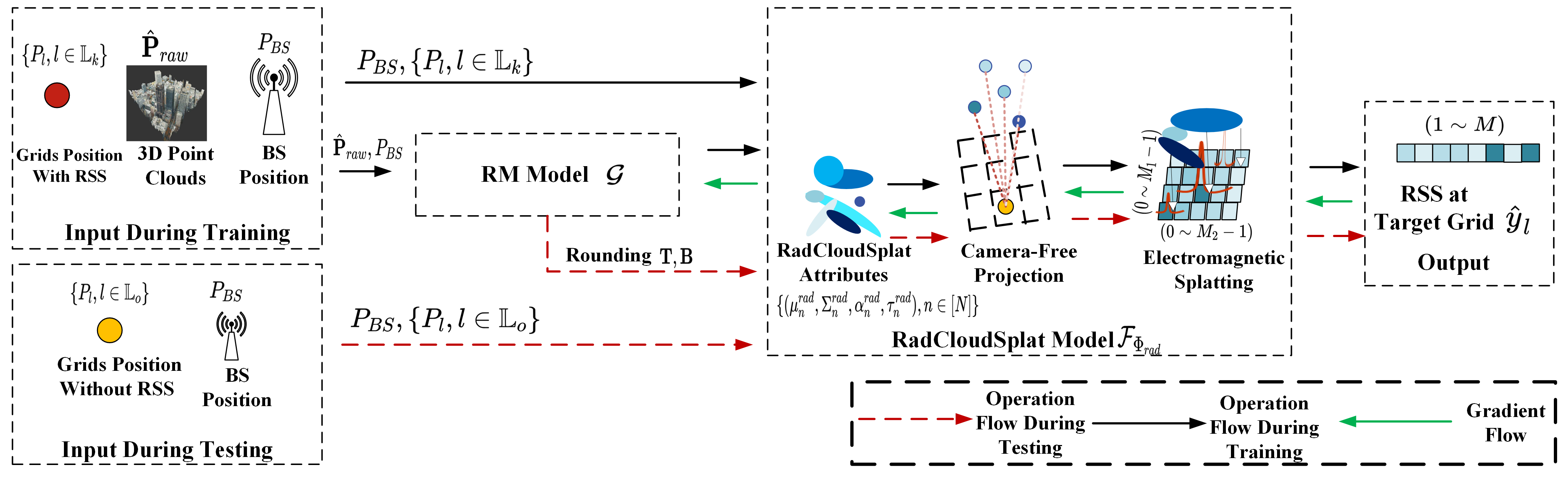}
    \caption{\textbf{An Overview of the RadCloudSplat Framework:}  The raw 3D point clouds are efficiently denoised and scaled down by the optimization-based \textit{RM} sub-model. This preprocessing model reparameterizes the positions of scatters for the RadCloudSplat framework. Once the scatterers' positions are identified, additional signal-related RadCloudSplat attributes are optimized to ensure accurate RSS predictions. This design encodes virtual scatterers as a set of 3D Gaussians, each embedding attenuation and signal characteristics. To project these 3D Gaussian representations onto the \textit{beamspace} for beam-wise RSS, a novel \textit{camera-free projection} technique is employed. Finally, the RSS values at the target grid are rendered using the \textit{electromagnetic splatting method}.}\label{fig:Framework}
    \vspace{-4mm}
\end{figure*}
\subsection{Relaxed-Mean Reparameterization  for Key Virtual Scatters Positions Extraction}\label{sec:relax}

\textbf{Initial formulation: }Identifying key virtual scatterers in RM can be initially expressed as a geometric problem. The scatterer positions \( \{\bm{\mu}_n^{\text{rad}}\}_{n=1}^N \) should represent a subset of the raw point cloud \( \hat{P}_{\text{raw}} \), while satisfying radio-propogation-related constraints. Formally, this can be expressed as
\begin{subequations}
\label{eq:optim1}
\begin{align}
    \text{Find }& \{\bm{\mu}_n^{\text{rad}}\}_{n=1}^N \in \mathbb{R}^{N \times 3}, \label{findobj} \\
     \text{s.t.}& \quad \hat{P}_N \subset \hat{P}_{\text{raw}},  \label{findc1} \\ 
    & \quad\| \{\bm{\mu}_n^{\text{rad}}\}_{n=1}^N - \hat{P}_N \|_2 \leq \epsilon_1, \label{findc2} \\&\quad \| \hat{P}_N - P_{\text{BS}} \|_2 \leq \epsilon_2. \label{findc3}
\end{align}
\end{subequations}
Here, \( \hat{P}_N \) is an intermediate subset of \( \hat{P}_{\text{raw}} \),  constraint (\ref{findc1})  ensures that the selected scatterers are derived from the original point cloud, preserving environmental features captured by the LiDAR sensor, constraint (\ref{findc2})  enforces that the final scatterers' positions \( \{\bm{\mu}_n^{\text{rad}}\}_{n=1}^N \) remain close to \( \hat{P}_N \), mitigating the impact of noise in the raw point cloud, and constraint (\ref{findc3}) prioritizes scatterers closer to the BS, as they contribute to dominant signal paths, such as LoS and strong NLoS reflections.

While this geometric formulation provides an intuitive starting point, it does not explicitly account for the relationship between scatterers' positions and RSS predictions. The ultimate goal of RM is to identify scatterers' positions \( \{\bm{\mu}_n^{\text{rad}}\}_{n=1}^N \) and other scatterers' parameters that minimize the error in predicting RSS values at target locations. To optimize for RSS prediction, we reformulate the problem as minimizing a loss function \( \ell(\cdot, \cdot) \) between the predicted RSS \( \hat{\bm{y}}_l \) and the ground truth RSS \( \bm{y}_l^{\text{GT}} \) collected at locations \( l \in \mathbb{L}_k \)
\begin{equation}
    \min_{\bm{\Phi}_{\text{rad}}} \ell \left( \hat{\bm{y}}_l, \bm{y}_l^{\text{GT}} \right), \quad \forall l \in \mathbb{L}_k,
\end{equation}
where \( \hat{\bm{y}}_l\) will be produced by  the RadCloudSplat rendering representing by (\ref{eq:3dgsRSS}) with  RadCloudSplat attributes $\bm{\Phi}_{\text{rad}} $ as will be illustrated in Section \ref{sec:Synthesis}.

\textbf{Relaxed Reparameterization:} To enable efficient selection of scatterers' positions, we reparameterize \( \bm{\mu}_n^{\text{rad}} \) using a trainable selection matrix \( \bm{T} \in \mathbb{R}^{N \times R} \) and bias term \( \bm{B} \in \mathbb{R}^{N \times 3} \) as
\begin{equation}
    \{\bm{\mu}_n^{\text{rad}}\}_{n=1}^N = \bm{T} \hat{P}_{\text{raw}} + \bm{B},
\end{equation}
where \( \bm{T} \) selects a subset of points from the raw point cloud \( \hat{P}_{\text{raw}} \), and \( \bm{B} \) refines these positions to mitigate noise. Thus, the intermediate subset $\hat{P}_{{N}}$ in (\ref{findc1}) can be regarded as $\hat{P}_{{N}}=\bm{T} \hat{P}_{\text{raw}}$. The selection matrix \( \bm{T} \) is constrained to the simplex space
\begin{equation}
    \bm{T} \in \mathbb{S} = \left\{ \bm{T} \in \mathbb{R}^{N \times R} \, \bigg| \, \bm T_{n,r} \geq 0, \, \sum_{r=1}^R \bm T_{n,r} = 1 \forall n \in [N] \right\}.
\end{equation}
To ensure sparsity and distinctiveness, we enforce a Maximum Element Constraint (MEC), which requires the largest element in each row of \( \bm{T} \) to dominate
\begin{equation}
    \max_r \bm T_{n, r} \gg \bm T_{n, r'}, \quad \forall n \in [N], \, r' \neq \arg\max_r \bm T_{n, r}.\label{eq:mec}
\end{equation}

After reparameterization, the final optimization problem is
\begin{subequations}
\label{eq:optim2}
\begin{align}
    \min_{\bm{T}, \bm{B}, \bm{\Phi}_{\text{rad}} \setminus \{\bm{\mu}_n^{\text{rad}}\}_{n=1}^N}& \ell \left( \hat{\bm{y}}_l, \bm{y}_l^{\text{GT}} \right) + \ell_{\text{reg}},\\
    \text{s.t.}&
   \quad \{\bm{\mu}_n^{\text{rad}}\}_{n=1}^N = \hat{P}_{{N}} + \bm{B},\label{eq:mu} \\
   &\quad \hat{P}_{{N}}=\bm{T} \hat{P}_{\text{raw}},\label{eq:mu1} \\
    &\quad \bm{T} \in \mathbb{S}.
    \end{align}
\end{subequations}
The regularization term \( \ell_{\text{reg}} \) can be derived by substituting (\ref{eq:mu}) and (\ref{eq:mu1}) into  (\ref{findc2}) and (\ref{findc3}), enforcing design principles such as proximity to the BS, noise mitigation, and sparsity in the selection matrix \( \bm{T} \) as
\begin{equation}
\begin{aligned}
    \ell_{\text{reg}} =& \lambda_1 \|\bm{T} \hat{P}_{\text{raw}} - P_{\text{BS}} \|_2 + \lambda_2 \|\bm{B}\|_2 \\
    &+ \lambda_{\text{MEC}} \|\bm{T} - \bm{1}\|_{\infty, 2} + \lambda_{\text{sparsity}} \|\bm{T}\|_1,
\end{aligned}
\end{equation}
where $\|\bm{T} - \bm{1}\|_{\infty, 2} = \sqrt{\sum_{n=1}^N \max_{r \in [R]} (\bm{T}_{n,r} - 1)^2}$. The \( \ell_{\infty, 2} \)-norm enforces \textit{row-wise dominance}, encouraging one dominant entry per row, while the \( \ell_1 \)-norm promotes \textit{global sparsity}. Together, these regularizers ensure that \( \bm{T} \) satisfies the MEC (\ref{eq:mec}) by encouraging distinct associations between the intermediate subset \( \hat{P}_{N} \) and the raw point cloud \( \hat{P}_{\text{raw}} \).



\subsection{RadCloudSplat for RSS Synthesis}
\label{sec:Synthesis}

After efficiently extracting the positions,$\{\bm{\mu}_n^{\text{rad}}\}_{n=1}^N$ of key virtual scatterers through the RM reparameterization model (Section \ref{sec:relax}), the next step in RadCloudSplat is to determine the remaining properties of the major virtual scatterers to accurately synthesize RSS values. These properties, referred to RadCloudSplat attributes, include attenuation signal  \(\{\bm{\alpha}_n^{\text{rad}} \}_{n=1}^N\), RSS-encoded SH coefficients \( \{\bm{\tau}_n^{\text{rad}}\}_{n=1}^N \), and the covariance matrices \( \{\bm{\Sigma}_n^{\text{rad}}\}_{n=1}^N \).
These attributes are optimized to align with the physical principles of wireless signal propagation, ensuring accurate RSS predictions.

\textbf{RadCloudSplat Attributes:}  The signal attenuation  \(\bm \alpha_n^{\text{rad}}\) models attenuation strength factor caused by the distance from the BS and corresponding phase shift. The direction vector from the BS to the \( n \)-th scatterer is defined as \( \bm{d}_n = {P}_{\text{BS}} - \bm{\mu}_n^{\text{rad}} \). To enhance spatial resolution, we encode the distance information using a position encoding function inspired by \cite{tancik2020fourier}
$    \gamma(\bm{d}_n) = \big[\sin(\pi \bm{d}_n), \cdots, \sin(\pi^V \bm{d}_n), \cos(\pi^V \bm{d}_n)\big],$
where \( V \) is the order of the encoding. A shallow multi-layer perceptron (MLP) is then applied to embed \( \gamma(\bm{d}_n) \)  to derive the complex attenuation signal with phase shift ${\bm \phi_n^{\text{rad}}}$
\begin{equation}
    \bm{\alpha}_n^{\text{rad}} =\textbf{MLP}_1(\gamma(\bm{d}_n); \bm{\Phi}_1)\odot e^{j\bm{\phi}^{\text{rad}}_n}, \quad \forall n \in [N],\label{eq:alpha}
\end{equation}
where \( \bm{\Phi}_1 \) represents the trainable parameters of the MLP, and each element of the phase shift ${\bm \phi_n^{\text{rad}}}$ is constrained to the interval $[-\pi,\pi]$ by the scaled hyperbolic tangent function. This formulation ensures that \(\bm \alpha_n^{\text{rad}} \) accurately models the scatterer's attenuation effect. 

To model view-dependent variations in RSS, we adapt SH functions, commonly used in visual tasks, to encode RSS \textit{beamspace} information. Specifically, we introduce RSS-encoded SH coefficients \( \bm{\tau}_n^{\text{rad}} \in \mathbb{R}^{M \times (k+1)^2} \), where \( M = M_1 \times M_2 \) is the total number of beams and \( k \) is the degree of the SH function. The covariance matrices \( \bm{\Sigma}_n^{\text{rad}} \) define the anisotropic spatial uncertainty of each scatterer.

With the RadCloudSplat attributes \(\bm{\mu}_n^{\text{rad}}\), \( \bm{\Sigma}_n^{\text{rad}} \), \(\bm{ \alpha}_n^{\text{rad}} \), and \( \bm{\tau}_n^{\text{rad}} \) defined, the next step in RadCloudSplat is the rendering process, which involves a camera-free projection of the scatterers onto the \textit{beamspace} and	synthesizing the beam-wise RSS values by \textit{electromagnetic splatting}.

\textbf{Camera-free Projection:}  A novel \textit{camera-free projection approach} is proposed to map the 3D virtual scatters onto the 2D plane of \textit{beamspace}, in Section ~\ref{sec:radioprop}. Unlike traditional 3D Gaussian splatting, which computes color values for pixels, RadCloudSplat aggregates contributions of scatterers to synthesize beam-wise RSS values. There are several key differences compared with the camera model in the optical 3DGS-based model: First, there are no intrinsic or extrinsic camera parameters. Second, although a closed-form physical projection exists for mapping spatial 3D to the 2D beam-wise RSS \textit{beamspace}, accurately determining the necessary parameters, e.g. antenna gain patterns and beamforming patterns \cite{johannsen2020single}, remains challenging in real-world scenarios since they are proprietary information owned by operators, and even inevitable errors always exist. To address this, we introduce two encoded-decoded structured MLPs for a camera-free projection that recovers the 2D Gaussian representation: the means $\bm{x} \in \mathbb{R}^2$ and precision, w.r.t. the inverse of variance, ${\bm \Lambda}_n \in \mathbb{R}^{2 \times 2}$
\begin{align}
    \bm{x}_n &= \textbf{MLP}_2({\bm \mu}^{\text{rad}}_n; \bm{\Phi}_2) \odot \bm{o}, \quad \forall n \in [N], \nonumber \\
    {\bm \Lambda}_n &= \textbf{MLP}_3({\bm \Sigma}^{\text{rad}}_n; \bm{\Phi}_3), \quad \forall n \in [N],
    \label{eq:2d}
\end{align}
where $\bm{\Phi}_2$ and $\bm{\Phi}_3$ are the trainable parameters of the means-projection and precision-projection functions. Here, $\bm{o} = [M_1 - 1, M_2 - 1]$ is a vector. The activation function of the last layer in the means-projection function $\textbf{MLP}_2(\cdot; \bm{\Phi}_2)$ is set as a Sigmoid function to constrain the $\bm{x}_n$ remains within the $M_1 \times M_2$ beam-wise RSS \textit{beamspace}, as shown in the top right corner of Fig.~\ref{fig:System Model}. To compute $\bm{w}_n \in \mathbb{R}^{M_1 \times M_2}$ of the 2D Gaussian splats over the beam-wise RSS \textit{beamspace}, given the precision and means from Eq.~(\ref{eq:2d}), the sampled indices of the \textit{beamspace} ${\bm{m}}(m_1, m_2) = [m_1-1, m_2-1] \in \mathbb{R}^2, \forall m_1 \in [M_1], \forall m_2 \in [M_2]$ can be used
{
\begin{align}
    \bm{w}_n(m_1,m_2) = \exp{\left(-\frac{t}{2} {\bm{g}_n(m_1,m_2)}^\text{T} {\bm \Lambda}_n \bm{g}_n(m_1,m_2) \right)},
    \label{eq:end}
\end{align}}
where ${ {\bm g_n}({m_1,m_2})}=\bm{m}({m_1,m_2})-\bm x_n, \forall m_1\in[M_1],\forall m_2\in[M_2]$ and $t$ is a scaling constant.

\textbf{Electromagnetic Splatting:} After the camera-free coordinate projection transformation, during rendering, each 2D Gaussian contributes to multiple RSS beams. Specifically, the depth ordering of the 3D Gaussian means, ${\bm \mu^{\text{rad}}_n, n \in [N]}$, relative to the target grid $P_l$, determines their contribution. These 2D Gaussians are sorted by ascending depth order. This sorting is represented by the permutation operation ${\bm \pi}_l: \bm \mu \to \Pi^l \bm\mu$, where $\Pi^l$ is a meaningful permutation matrix denoting the sorting operation as a one-to-one mapping $n \to {\bm \pi}_l(n), n \in [N]$. The value of each RSS beam is the cumulative contribution of all sorted 2D Gaussians, as illustrated in the \textit{electromagnetic splatting module} in Fig.~\ref{fig:Framework}. To reconstruct the beam-wise RSS for the target $l$-th grid, and following Eq.~(\ref{discretizing}), the \textit{electromagnetic splatting module} integrates the attributes of the $n$-th scatter as 
\begin{align} \Delta \hat{\bm\beta}_n \odot e^{j\Delta\hat{\bm{\varphi}}_n} = \bm{c}^{\text{rad}}_{{\bm \pi}_l(n)} \odot\bm{\alpha}^{\text{att}}_{{\bm \pi}_l(n)}\odot \textbf{vec}(\bm{w}_{{\bm \pi}_l(n)}), \end{align} 
where $\textbf{vec}(\cdot)$ denotes the matrix vectorization operation, and $\bm{c}^{\text{rad}}_{{\bm \pi}_l(n)} \in \mathbb{C}^M$ represents the view-dependent radio signal as
\begin{align} \bm{c}^{\text{rad}}_{{\bm \pi}_l(n)} = \bm{sh}(\bm{\tau}^{\text{rad}}_{{\bm \pi}_l(n)}, \bm{v}^{\text{rad}}_{{\bm \pi}_l(n)}, k) \odot e^{j\bm{\varphi}^{\text{rad}}_{{\bm \pi}_l(n)}}, \quad \forall n \in [N], \label{eq:begin} \end{align}
where the view direction is $\bm{v}^{\text{rad}}_{{\bm \pi}_l(n)} = \frac{P_l - \bm{\mu}^{\text{rad}}_{{\bm \pi}_l(n)}}{|P_l - \bm{\mu}^{\text{rad}}_{{\bm \pi}_l(n)}|2}$, and $\bm{\varphi}^{\text{rad}}_{{\bm \pi}_l(n)} \in \mathbb{R}^M$ denotes the signal phase. The final attenuation signal $\bm{\alpha}^{\text{att}}_{{\bm \pi}_l(n)}$ accounts for the accumulated attenuation effects on the ${{\bm \pi}_l(n)}$-th scatter, expressed as
\begin{align} \bm{\alpha}^{\text{att}}_{{\bm \pi}_l(n)} = \bm{\hat{\alpha}}^{\text{rad}}_n \odot \prod_{i=1}^{n-1} (1 - \bm{\hat{\alpha}}^{\text{rad}}_i) \in \mathbb{C}^M, \end{align} 
with the sorted attenuation signal $\hat{\bm{\alpha}}^{\text{rad}}_n = \bm{\alpha}^{\text{rad}}_{{\bm \pi}_l(n)}$, $\forall n \in [N]$. These formulations assume that each scatter acts as a virtual transmitter, retransmitting a new complex signal modified by an accumulated complex attenuation factor. Using the RSS-related Eq.~(\ref{eq:rss}), the final blended RSS $\hat{\bm{y}}_l$ is computed as \begin{align} \hat{\bm{y}}_l = P\left|\sum_{n=1}^{N}\Delta \hat{\bm\beta}_n \odot e^{j\Delta\hat{\bm \varphi}_n}\right|^2, \end{align} where the transmit power $P$ is usually known in advance, and the term $\Delta\hat{\bm \varphi}_n$ captures the combined effect of the phase shift ${\bm \phi}_{n}^{\text{rad}}$ in Eq.~(\ref{eq:alpha}) and signal phase $\bm{\varphi}^{\text{rad}}_{{\bm \pi}_l(n)}$ in Eq.~(\ref{eq:begin}).}

\subsection{Optimizing RadCloudSplat}\label{sec:loss}

Summarizing Sections~\ref{sec:relax} and~\ref{sec:Synthesis}, {the trainable parameters of RadCloudSplat are from the reparameterization of RadCloudSplat attributes, the camera-free projection, and the electromagnetic splatting listed as 
\[
\bm{\Gamma}_{\text{rad}} = \{\bm{T},\bm{ B}, \bm{\Phi}_1, \bm{\Phi}_2, \bm{\Phi}_3, \{\bm{\tau}_n^{\text{rad}}, \bm{\Sigma}_n^{\text{rad}}, \bm{\phi}_n^{\text{rad}}, \bm{\varphi}_n^{\text{rad}}\}_{n=1}^N\}.
\]
Combining with the optimization formulation of \( \bm{T} \) and \( \bm{B} \) in (\ref{eq:optim2}), we train the RadCloudSplat models using the following loss with mini-batch gradient descent
\begin{equation}
    \ell_{\text{final}}(\bm\Gamma_{\text{rad}}) = \ell\left(\hat{\bm y}_l, \bm y_l^{\text{GT}}\right) + \ell_{\text{reg}},\label{finalloss}
\end{equation}
where \( \ell \) is a composite loss combining mean squared error (MSE) and total variation (TV) losses
\begin{equation}
\ell =  \underbrace{\frac{1}{|\mathbb{D}|} \sum_{l \in \mathbb{D}} (\hat{\bm y}_l - \bm{y}^{GT}_l)^2}_{\text{MSE Loss}} + \lambda_{\text{TV}} \cdot \underbrace{\frac{1}{|\mathbb{D}|} \sum_{l \in \mathbb{D}} \| \hat{\bm y}_l - \hat{\bm y}_{\text{KNN}(l)} \|_1}_{\text{TV Loss}}.
\end{equation}
Here, \(\lambda_{\text{TV}}\) controls the TV loss weight and \(\mathbb{D}\) is the mini-batch sample set with \(|\mathbb{D}|\) samples. \(\hat{\bm y}_{\text{KNN}(l)}\) denotes the predicted RSS value at the K nearest neighboring grid locations to the \(l\)-th grid position within \(\mathbb{D}\). By balancing accuracy, smoothness, and regularization, the final ternary area-by-area loss (\ref{finalloss}) provides a robust framework for RSS radiomap extrapolation, ensuring precise and spatially coherent predictions.} 

{Given training set \( \mathbb{L}_k \) and testing set \( \mathbb{L}_o \), in addition to the design of the loss function, a novel training scheme, termed the recursive fine-tuning scheme, is proposed to further enhance performance. Observing that data points within the grids \( \mathbb{L}_o \) closer to the boundary, which separates the grids \( \mathbb{L}_k \) and the grids \( \mathbb{L}_o \), are easier to recover, this scheme leverages the nature of the extrapolation task within spatial data to progressively recover data near the boundary \(\mathbb{ L}_D \). The recursive fine-tuning scheme for outdoor scenes consists of three key steps:  \begin{itemize}
    \item {\bf Step 1. Training on the Existing Training Set:} Train the RadCloudSplat model by minimizing the final loss function (\ref{finalloss}) on grids \(\mathbb{ L}_k \), where RSS is assumed known, for predefined \( K \) iterations. 
    \item  {\bf Step 2.  Predicting RSS Near the Boundary:} Identify grids \(\mathbb{ L}_D \) within \( D \) units of distance near boundaries from \( \mathbb{L}_o \), and predict the RSSs at these grids using the trained RadCloudSplat model by step 1.
    \item {\bf Step 3. Constructing a New Training and Testing Set:} Incorporate the predicted RSS values and their corresponding grids \( \mathbb{L}_D \) into the existing training set \( \mathbb{L}_k \) and remove them from the testing set \( \mathbb{L}_o \): $
\mathbb{L}_k = \{l \,|\, l \in \mathbb{L}_k \cup \mathbb{L}_D\}, \quad \mathbb{L}_o = \{l \,|\, l \in \mathbb{L}_o \setminus \mathbb{L}_D\}.$
\end{itemize}
These three steps are repeated iteratively until the RSS values for all grids in \(\mathbb{ L}_o \) are covered by \(\mathbb{ L}_k \). Then, the RadCloudSplat model with trained parameters is selected as the final model to evaluate on the original testing set \( \mathbb{L}_o \). Specifically, in the \textit{testing phase}, instead of directly using \(\bm T \) produced in the \textit{training phase}, we round it by selecting the maximum element in each row and setting other elements to zero, ensuring the rounded \( \bm T \) being binary selection matrix \({\bm T} \in \mathbb{B}^{N \times R} \), where \( \sum_{r=1}^R {\bm T}_{n,r} = 1, \forall n \in [N] \). Then we use $\{{\bm \mu}^{\text{rad}}_n\}_{n=1}^{N}$ determined by these for RadCloudSplat.}
\subsection{Summary}
RadCloudSplat is grounded in principles from radio propagation physics and optimization theory. On the physics side, we treat each identified 3D virtual scatterers as secondary sources that re-transmit radio waves with phase and amplitude characteristics, consistent with standard multipath models; On the optimization theory side, the RM scheme adopts Boolean relaxation to relax the undifferentiable binary selection matrix to the differentiable matrix on a simplex to ensure the convergence of the optimization of the position of the 3D Gaussians. 
\section{Synthetic Data Evaluation}
To better understand the performance of RadCloudSplat in representing virtual scatters with several key radio-related attributes versus its performance in extrapolating RSS radiomap in unseen locations, we first evaluate the RadCloudSplat relying on a limited set of measurements on synthetic datasets.

\textbf{Synthetic Datasets:} We use the synthetic data sets based on City A in China with $5,963,068$ point clouds captured by LiDAR sensor to describe the outdoor complex environment. Fig.~\ref{fig:Task} shows the layout geometry of the outdoor environment, whose size is around $1000\times1200\times 110.5m^3$, and the spatial distribution of outdoor RSS data from central $6,310$ grids (each of $10\times10m^2$), all of which are measured at the same height $1m$ by Ray Tracing method. There is only one main BS in this unit coverage area. To emulate the signal interference encountered in real-world scenarios, we incorporate $3$ dB noise into the data. Moreover, we separate the total measurements into the training set with $|\mathbb L_k|=5,566$ red grids and the testing set with only $|\mathbb L_o|=744$ yellow grids to evaluate.
\subsection{Experiments Design and Implementation}
\textbf{Benchmarks.} We compare RadCloudSplat with several baselines\footnote{The projection model in the previous WRF-GS method differently projects the virtual transmitters, represented by 3D Gaussians, onto the perception plane of the receiver antenna array. However, this design inherently limits its ability to model signal propagation to the RSS {\em beamspace}, particularly in outdoor radiomap reconstruction tasks where multi-beam effects and complex propagation paths are prevalent.} :
\begin{itemize}
    \item \textbf{Kriging Method:} This approach is the typical model-based method for spatial data, which is also beneficial for radiomap-related interpolation tasks.
    \item \textbf{Variational Autoencoder (VAE):}VAE is commonly used to generate new data samples by mapping observed data to a continuous latent space. Here, we use VAE to predict the spatial RSSs at target grids.
    \item \textbf{$\text{NeRF}^2$:} $\text{NeRF}^2$ is an SOTA learning-based method for spatial spectrum synthesis using the NeRF technique. After training, $\text{NeRF}^2$ can predict the spatial  RSSs at arbitrary target grids' positions.
    \item \textbf{$\text{NeWRF}$:} Similar to $\text{NeRF}^2$, $\text{NeWRF}$ can also predict the spatial RSSs at arbitrary target grids' positions.
    \item \textbf{RadCloudSplat with Physical Projection (RadCloudSplat w/ PP):} To assess the effectiveness of our MLP-based camera-free projection approach, we implement a RadCloudSplat variant baseline that utilizes a physical projection method. Specifically, this baseline assumes a predefined uniform planar array (UPA) at the BS and employs a standard DFT codebook\cite{love2005limited} for beamforming. 

\item \textbf{RadCloudSplat without Recursive Fine-tuning (RadCloudSplat w/o RF):} To evaluate the impact of our recursive fine-tuning strategy, we introduce a RadCloudSplat variant that omits the recursive fine-tuning process.

\end{itemize}

\textbf{Details of Implementation.} We implement RadCloudSplat in Pytorch. To process dense point clouds, we first extract edge points, which are typically defined by significant changes in surface normal vectors \cite{demarsin2007detection}, such as boundaries of obstacles or regions with differing properties, and are more likely to represent virtual scatterers. Accurately identifying these edge points enhances model quality and reduces time costs. This results in a subset of points, maintaining the structural integrity of the original cloud \cite{qi2017pointnet++}. The RadCloudSplat, with $N=2,000$ points, can be trained using the Adam optimizer \cite{kingma2014adam} and the area-by-area loss function in Eq.~(\ref{finalloss}). For $\text{NeRF}^2$ and NeWRF, we adapt the open-source codes \cite{zhao2023nerf2,lunewrf} for the RSSI prediction task to our RSS radiomap extrapolation task. All baselines are retrained on our datasets and evaluated on the same test sets.
\subsection{Synthetic RSS Radiomap Extrapolation Results}
We first evaluate how efficiently and accurately our proposed RadCloudSplat predict RSS compared to three baselines on the inference cost time\footnote{All experiments are carried out on NVIDIA A30.} and the metric Mean Absolute Error (MAE), whose lower values indicate greater similarity between predicted and ground truth samples, defined as
{\begin{align}
    \frac{1}{M}\frac{1}{|\mathbb L_o|}\sum_{m=1}^M \sum_{l\in\mathbb L_o}^{} |10\log(\hat{\bm{y}}_{m,l}) - 10\log({\bm{y}}_{m,l}^{GT})|,
\end{align}}
\hspace{-3pt}where $|\mathbb L_o|$ is the total number of grids in the test dataset and $\hat{\bm{y}}_l$ denotes the predicted RSS from different models.


\begin{table}[ht]
\vspace{-3mm}
    \centering
    \caption{Performance Comparison on Synthetic Dataset}
    \label{tab:shanghai}
   \begin{tabular}{lccc}
        \toprule
        \textbf{Method} & \textbf{MAE (dB)} & \textbf{IT (s)}&\textbf{TT (s)}\\
        \midrule
        Kridging       & 11.307& 0.010&-\\
        VAE       & 10.306 & 0.007&1.348\\
        $\text{NeWRF}$       & 10.051 & 0.032&2219.258\\
        $\text{NeRF}^2$       & 9.867 & 0.020&2210.355\\
        RadCloudSplat w/ PP  & 12.218 & 0.010&26.412 \\
        RadCloudSplat w/o RF  & 7.967 & 0.004& 1.652\\
        RadCloudSplat  & 7.564 & 0.004& 1.652\\
        
        \bottomrule
    \end{tabular}
\end{table}
TABLE~\ref{tab:shanghai} compares MAE metric values and inference/training time (IT and TT) costs\footnote{ Inference time cost refers to the average duration for a model to complete one inference cycle on a test data point. Training time represents the duration required for an algorithm to complete one full training cycle (i.e., one epoch) on a given training dataset.} on the  City A synthetic dataset. Among all compared methods, RadCloudSplat achieves the best performance, producing predictions most consistent with the ground truth and yielding a MAE of only $7.564$ dB. In contrast, all baseline methods—excluding RadCloudSplat variants—produce MAE values exceeding $9$ dB, highlighting the superior accuracy of our proposed model. The slightly inferior performance of NeRF-based models is due to challenges in locating the spatial positions of virtual scatterers along propagation paths without using any environmental features. The results also report the rendering time required to synthesize a single RSS value at a target grid. RadCloudSplat demonstrates significantly faster inference, rendering one sample in only $0.004$ seconds, compared to $\text{NeRF}^2$ and $\text{NeWRF}$, which require $0.02$ and $0.03$ seconds, respectively—making RadCloudSplat approximately five times faster. Furthermore, training time comparisons indicate that RadCloudSplat achieves superior training efficiency, requiring less time per epoch than baseline models. Thus, RadCloudSplat demonstrates superior computational efficiency and is suitable for latency-sensitive applications, whereas the NeRF-based model faces time complexity challenges due to the need to search the whole angular space before rendering.

Further ablation studies comparing RadCloudSplat (MAE $7.564$ dB) with its variant RadCloudSplat w/o RF (MAE $7.967$ dB) demonstrate that the extrapolation capability improves significantly when the recursive fine-tuning strategy is applied. Additionally, comparisons with the RadCloudSplat w/ PP variant (MAE $12.218$ dB) clearly show that our MLP-based projection not only achieves better performance but also improves robustness in the absence of operator-specific projection parameters since the physical projection model depends on parameters such as the antenna gain patterns and beamforming patterns, which are proprietary and inaccessible in real-world settings.



\subsection{Impact of Extracted Virtual Scatters}\label{sec:ablation1}
To further investigate the impact of extracted virtual scatterers on the RSS extrapolation task, we conduct additional ablation experiments with three baselines, all of which are variants of RadCloudSplat: {1) In the \textit{Random-Points-Opt} model, $N$ points are randomly sampled from the $R$ raw point clouds $\hat{P}_{\text{raw}}$ using the farthest point sampling method \cite{moenning2003fast}. These sampled points serve as the initial means of the key virtual scatterers $\{\bm{\mu}^{\text{rad}}_n\}_{n=1}^N$, instead of employing the RM scheme to reparameterize them as in RadCloudSplat. Notably, the means $\{\bm{\mu}^{\text{rad}}_n\}_{n=1}^N$ in \textit{Random-Points-Opt} are directly optimized during training alongside other parameters $\bm{\Phi}_{\text{rad}} \setminus \{\bm{\mu}^{\text{rad}}_n\}_{n=1}^N$. 2) Different from the \textit{Random-Points-Opt} model, the \textit{Random-Initial-Opt} model initializes the means of key virtual scatters $\{\bm{\mu}_n^{\text{rad}}\}_{n=1}^N$ from a uniform distribution within the coverage area without using raw point cloud data $\hat{P}_{\text{raw}}$. 3) The \textit{Random-Initial-Fixed} model follows the same initialization as \textit{Random-Initial-Opt} for the means $\{\bm{\mu}_n^{\text{rad}}\}_{n=1}^N$ of key virtual scatters. However, unlike in \textit{Random-Initial-Opt}, these means remain fixed during training, and only the other parameters $\bm{\Phi}_{\text{rad}} \setminus \{\bm{\mu}_n^{\text{rad}}\}_{n=1}^N$ are optimized.}

\begin{table}[ht]
    \centering
   \caption{Ablation Experiments on Synthetic Dataset}
\label{tab:1}
   \begin{tabular}{lcc}
        \toprule
        \textbf{Method} & \textbf{MAE (dB)}\\
        \midrule
     RadCloudSplat& 7.564\\
Random-Points-Opt   & 8.167 \\
Random-Initial-Opt    & 10.463        \\
Random-Initial-Fixed& 53.234\\
        \bottomrule
    \end{tabular}
\end{table}

{The performance results are summarized in TABLE~\ref{tab:1}, leading to three key conclusions: 1) Comparing \textit{Random-Initial-Fixed} and \textit{Random-Initial-Opt} highlights the necessity of optimizing the means of key virtual scatters $\{\bm{\mu}_n^{\text{rad}}\}_{n=1}^N$ in the RadCloudSplat model to enhance its expressiveness, demonstrating the importance of introducing bias \(\bm B\) to add flexibility. 2) Comparing \textit{Random-Points-Opt} and \textit{Random-Initial-Opt} confirms that point cloud modality data significantly improve the RSS radiomap extrapolation task. 3) Comparing \textit{RadCloudSplat} and \textit{Random-Points-Opt} demonstrates that raw point clouds $\hat{P}_{\text{raw}}$, despite their noise, can effectively guide the selection of means of key virtual scatterers via the novel and efficient RM reparameterization model, outperforming random selection methods.}

\section{Real-world Data Evaluation}
To address the concern regarding robustness to realistic noise and scaling, we evaluated our model, RadCloudSplat, using data collected via a outdoor measurement campaign carried out using the drive test in real-world live networks in City B in China. This dataset includes inherent sensor noise, GPS inaccuracies, and environmental complexity—conditions that closely reflect practical deployment challenges. As a supplementary evaluation, we add a physics-driven baseline \textit{Ray Tracing} to the Benchmarks on the real-world dataset. Our proposed model has been evaluated on this dataset to demonstrate its practical applicability and performance.
\begin{figure}
\begin{center}
\centerline{\includegraphics[width=0.7\columnwidth]{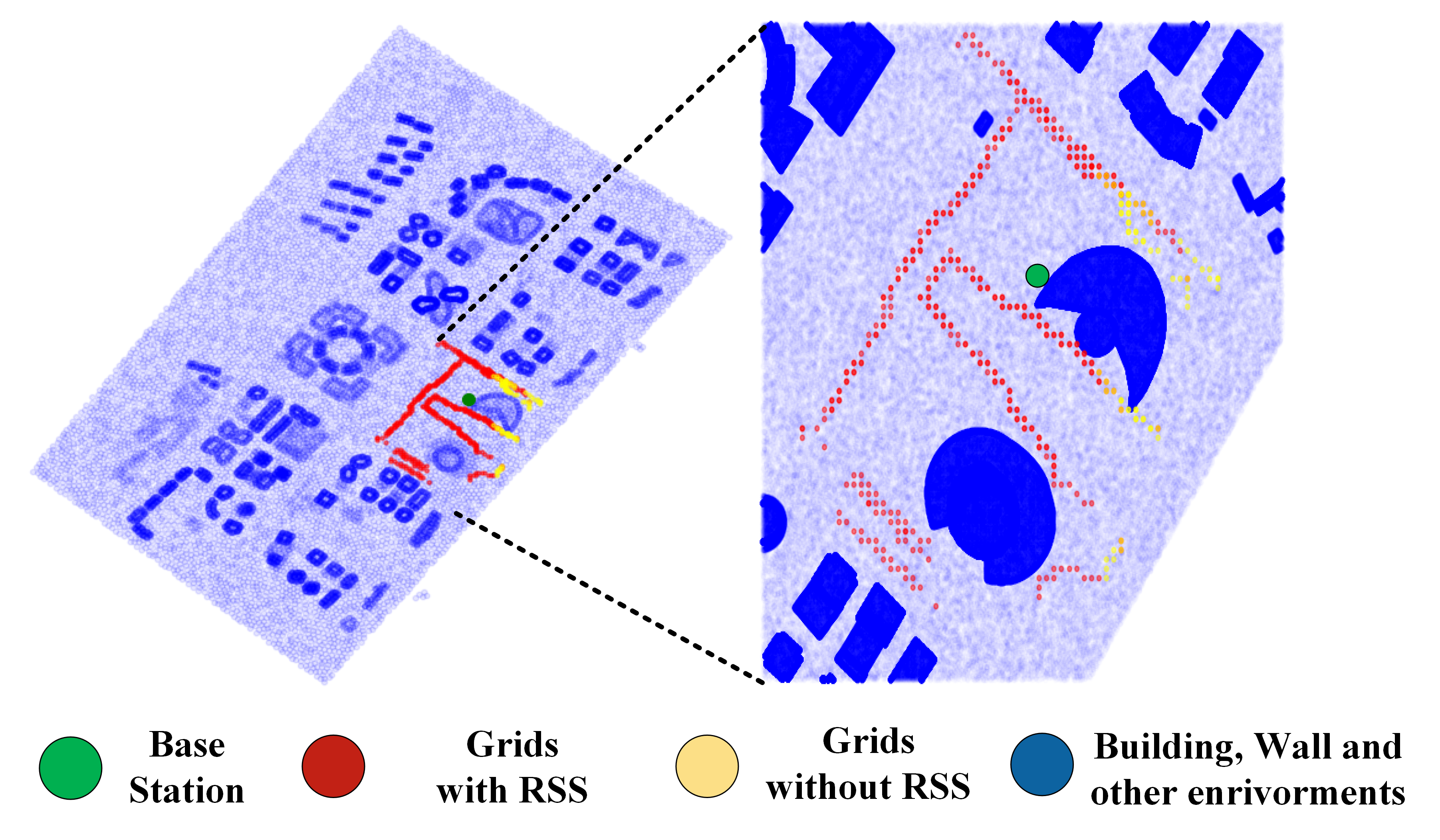}}
\caption{Illustration of City B's 3D Outdoor Environment.}
\label{hangzhou}
\end{center}
\vspace{-10mm}
\end{figure}

\textbf{Real-world Datasets:} The real-world datasets are collected from City B in China, consisting of $8,711,005$ LiDAR-captured point clouds to model obstacles, as shown in the zoomed view in Fig.~\ref{hangzhou}. RSSs data are collected from $492$ grids (\(10 \times 10 \, m^2\)) within an area containing a single main BS. Then the dataset is divided into $|\mathbb L_k|=417$ red grids for training and $|\mathbb L_o|=75$ yellow grids for testing.
\subsection{Real-world RSS Radiomap Extrapolation Results}
\begin{table}[ht]
\vspace{-4mm}
    \centering
    \caption{Performance Comparison on Real-world Dataset}
    \label{tab:hangzhou}
   \begin{tabular}{lccc}
        \toprule
        \textbf{Method} & \textbf{MAE (dB)} & \textbf{IT (s)}& \textbf{TT (s)}\\
        \midrule
        Ray Tracing       & 13.108 & 10.446& -\\
        Kriging       & 9.656 & 0.183& -\\
        VAE       & 8.728 & 0.011& 1.147\\
        $\text{NeWRF}$       & 7.806 & 0.563& 167.061\\
        $\text{NeRF}^2$       & 7.313 & 0.585& 165.524\\
        RadCloudSplat w/ PP  & 11.894 & 0.022& 2.964 \\
        RadCloudSplat w/o RF  & 7.121 & 0.018& 1.094 \\
        RadCloudSplat  & 7.035 & 0.018& 1.094 \\
        \bottomrule
    \end{tabular}
    \vspace{-4mm}
\end{table}

TABLE~\ref{tab:hangzhou} presents the prediction performance and computational efficiency of the RadCloudSplat framework for the real-world outdoor RSS radiomap extrapolation task. The results clearly demonstrate that the proposed RadCloudSplat framework outperforms other baseline methods while maintaining substantial computational efficiency, making it highly suitable for practical applications. 

\begin{figure*}[t]
\centering
   \centering
\includegraphics[width=0.7\textwidth]{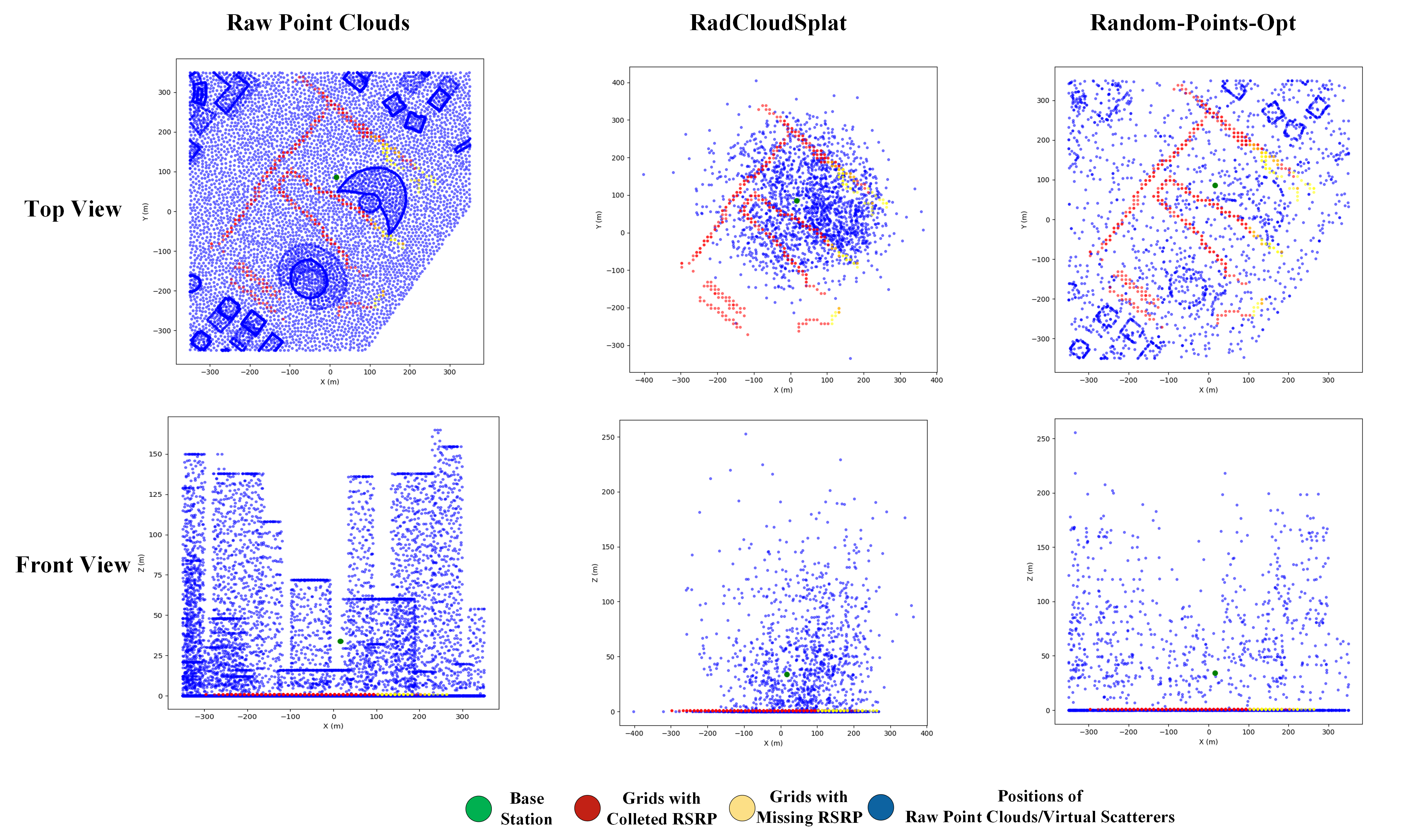}
    \caption{Visualization of Raw Point Clouds and Extracted Virtual Scatterers in City B}\label{fig:comp}
    \vspace{-2mm}
\end{figure*}

\subsection{Impact of RM Scheme}

In particular, with the aid of point cloud modality data and the white-box RM scheme introduced in Section~\ref{sec:relax}, the proposed RadCloudSplat model is capable of representing virtual scatterers and producing meaningful radiomap reconstructions. 

To qualitatively assess whether the positions of virtual scatterers fitted by the RM scheme correspond to physically plausible values, we visualize the raw point clouds of the City B dataset alongside the scatterer locations estimated by RadCloudSplat and its ablated variant, \textit{Random-Points-Opt}, as shown in Fig.~\ref{fig:comp}. This comparison illustrates that RadCloudSplat yields scatterer positions that align more coherently with the BS. To further support this observation quantitatively, we compute and compare the coverage percentage of scatterers near the BS in Fig.~\ref{fig:dig}, demonstrating that RadCloudSplat achieves superior proximity alignment to the BS relative to its variant.



\begin{figure}
\begin{center}
\vspace{-5mm}
\centerline{\includegraphics[width=0.5\columnwidth]{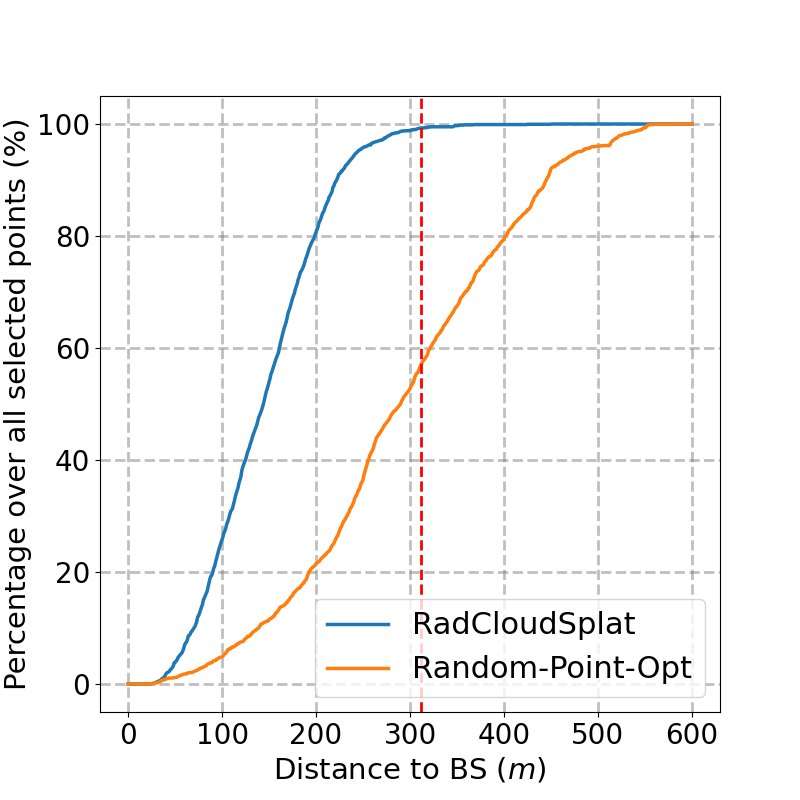}}
\caption{Illustration of Coverage Percentage Comparison.}
\label{fig:dig}
\end{center}
\vspace{-10mm}
\end{figure}

\begin{table}[ht]
\vspace{-4mm}
    \centering
   \caption{Ablation Experiments on Real-world Dataset}
\label{tab:4}
   \begin{tabular}{lcc}
        \toprule
        \textbf{Method} & \textbf{MAE (dB)}\\
        \midrule
     RadCloudSplat& 7.035\\
RadCloudSplat w/o Bias   & 8.434 \\
        \bottomrule
    \end{tabular}
\vspace{-2mm}
\end{table}



Additionally, we introduce a novel ablation study with a baseline named \textit{RadCloudSplat Without Bias Term (RadCloudSplat w/o Bias)}, where the scatterer positions are directly selected from raw point cloud data without any offset. Although this method yields zero mean error relative to the raw point cloud's positions, it results in significantly degraded performance in radiomap reconstruction, as shown in TABLE~\ref{tab:4}.

These findings confirm that the RM scheme effectively balances the noisy raw point cloud data with learned, task-specific corrections, enabling RadCloudSplat to achieve more accurate and physically plausible radiomap reconstructions.
\section{Conclusion}
In this work, we first extended 3DGS to the radio frequency domain, leveraging camera-free RadCloudSplat to extrapolate RSSs with high accuracy from sparse measurements in an outdoor environment. By efficiently selecting the means of key virtual scatterers from dense point clouds aided by the RM scheme, the model captured intricate multi-path propagation characteristics. Experiments and analysis validated the effectiveness of these scatterers, advancing the state-of-the-art in wireless network modeling and extrapolation performance and highlighting the transformative potential of integrating advanced 3D modeling techniques with wireless propagation analysis for next-generation applications in the radio domain. 

\newpage
\bibliographystyle{IEEEtran}
\bibliography{refs}

\end{document}